\pdfoutput=1

\documentclass[article,10pt,onecolumn,longbibliography]{revtex4-1}
\usepackage{microtype}
\usepackage{graphicx}
\usepackage[colorlinks,linkcolor=blue,urlcolor=blue,citecolor=blue]{hyperref}

\usepackage{caption}
\usepackage{subcaption}
\usepackage{mathptmx}
\usepackage{times}
\usepackage{epsfig,amsopn}
\usepackage{graphicx}
\usepackage{bm,amsmath,amssymb}
\usepackage{natbib}
\usepackage{braket}
\usepackage{float}
\usepackage{enumerate}
\usepackage{hyperref}
\allowdisplaybreaks[1]

\begin{document}
\def\be{\begin{equation}}
\def\ee{\end{equation}}
\def\bea{\begin{eqnarray}}
\def\eea{\end{eqnarray}}
\def\f{\frac}
\def\l{\label}
\def\nn{\nonumber}

\definecolor{dgreen}{rgb}{0,0.7,0}
\def\redw#1{{\color{red} #1}}
\def\green#1{{\color{dgreen} #1}}
\def\blue#1{{\color{blue} #1}}
\def\brown#1{{\color{brown} #1}}

\newcommand{\eref}[1]{~(\ref{#1})}%
\newcommand{\Eref}[1]{Equation~(\ref{#1})}%
\newcommand{\fref}[1]{Fig.~\ref{#1}} %
\newcommand{\Fref}[1]{Figure~\ref{#1}}%
\newcommand{\sref}[1]{Sec.~\ref{#1}}%
\newcommand{\Sref}[1]{Section~\ref{#1}}%
\newcommand{\aref}[1]{Appendix~\ref{#1}}%
\newcommand{\sgn}[1]{\mathrm{sgn}({#1})}%
\newcommand{\erfc}{\mathrm{erfc}}%
\newcommand{\Erf}{\mathrm{erf}}%

\title{Integral Fluctuation Theorems for Stochastic Resetting Systems}

\author{Arnab Pal and Saar Rahav}
\affiliation{Schulich Faculty of Chemistry, Technion-Israel Institute of Technology, Haifa 32000, Israel}
\date{\today}

\begin{abstract}
We study the stochastic thermodynamics of resetting systems. Violation
of microreversibility means that the well known derivations of fluctuations theorems
break down for dynamics with resetting. Despite that we show that stochastic resetting
systems satisfy two integral fluctuation theorems. The first is the Hatano-Sasa relation
describing the transition between two steady states. The second integral fluctuation theorem
involves a functional that includes both dynamical and thermodynamic contributions. We find that the second
law-like inequality found by Fuchs {\it et al.} for resetting systems [{\it EPL}, {\bf113}, (2016)] can be recovered from this integral
fluctuation theorem with the help of Jensen's inequality.
\end{abstract}


\maketitle

\section{Introduction}
Dynamics with resetting, where a system is intermittently returned to a predetermined state, has been
fascinating researchers from many fields and disciplines
\cite{EM11a,EM11b,KMSS14,EM14,Pal14,SabhapanditTouchette15,MSS15,Eule2016,RoldanGupta2017}.
Indeed, dynamical systems with resetting
have been employed as models for
diverse situations such as searching for lost possessions, foraging for food in the wild, stochastic phenotype switching,
optimal search algorithms,
and random catastrophic events \cite{BMS13,Majumdar:2005,Redner,First-Passage-Book,Benichou2011RMP}.
Part of the interest is due to the neat mathematical structure of resetting, but most of the interest is due to
the usefulness of resetting in search problems
\cite{EM11a,EM11b,KMSS14,Shamik16,SR2014,SR2016,AAM16,PalShlomi2016}.
It is now well understood that the inclusion of resetting can drastically affect the distribution of search times.
Consider a particle that diffuses until it reaches a target position for the first time.
Naively one would think that adding resetting to the system's dynamics is unlikely to be helpful, since
some resetting events will occur when the particle already is near the target, and will therefore be detrimental.
This naive intuition is often wrong. Resetting can be quite helpful in search problems in which a particle can diffuse
far away from the target in the wrong direction. Resetting prevents such realizations from occurring, thereby removing
realizations which take an exceedingly large times to reach their target \cite{EM11a,EM11b,SR2016,AAM16,PalShlomi2016}.

One of the most fundamental characteristics of stochastic resetting systems is that they are
inherently out of thermal equilibrium.
Consider a Brownian particle diffusing in some potential landscape. In the absence of a non-conservative force
(or boundary conditions that couple the system to imbalanced reservoirs)
the system will relax to an equilibrium state, in which the probability to find the system at $x$
is given by the Boltzmann distribution. Resetting can be added to the dynamics by mandating that at each time step $dt$
the system has probability $r dt$ to be reset to a preselected position $x_r$. If a reset has not happened the system
simply continues to diffuse. The resetting step is clearly unidirectional, since the dynamics does not include
anti-resetting transitions (i.e. the time-reversal of resetting events).
This simple observation means that stochastic dynamics with resetting can not satisfy detailed balance, and the system must therefore relax to
a non-equilibrium steady state \cite{Pal14}. It has been noticed that
the dynamics of relaxation to the steady-state can be quite unusual
in the presence of resetting
\cite{MSS15}. With time, an inner core region near the resetting point relaxes to the steady state, while the outer region
is still transient, and the location of the boundary separating them grows as a power law \cite{MSS15}.

While the dynamics of stochastic resetting systems was studied extensively, the thermodynamic
interpretation of resetting was largely overlooked. The first, and to the best of our knowledge the only
paper to deal with this question was published fairly recently \cite{Seifert2016}.
Fuchs {\it et. al.}  used the theory of stochastic thermodynamics in order to
give a consistent thermodynamic interpretation to resetting.
In particular,
the authors of Ref. \cite{Seifert2016} identified the entropy change and work due to a resetting event eventually deriving
the first and second law of thermodynamics in the presence of resetting. The change in the system's
entropy during a reset was interpreted as the difference between the information created and erased in this step, making an interesting
connection between resetting and the thermodynamics of information \cite{Parrondo2015}.

The theory of stochastic thermodynamics was developed to extend
thermodynamics into the realm of small out-of-equilibrium systems \cite{Sekimoto-book,Seifert:2012,Jarzynski:11}.
The most important concept underlying the theory is the ability of assigning a meaningful thermodynamic interpretation
to a single realization of a process. This in turn allows one to study distributions of thermodynamic quantities, such as heat or work
\cite{VanZon-Cohen,PalSanjib}.
The development of stochastic thermodynamics was largely motivated by the realization that such distributions satisfy
a set of results known as
fluctuation theorems \cite{Evans1993,Gallavotti1995,Kurchan1998,Lebowitz1999,Seifert2005,Jarzynski1997,Crooks1998}.
Fluctuation theorems (FT) can be viewed as replacing the inequality of the second law by an equality expressed as an
exponentially weighted average over distributions of thermodynamic variables. These celebrated
results are a rare example of general laws that hold even for far from equilibrium processes. Their discovery spurred an
extensive research effort focused on out-of-equilibrium systems and processes \cite{Seifert:2012,Jarzynski:11}.

It is only natural to ask whether stochastic resetting systems satisfy any fluctuation theorems. The question
is non trivial since the resetting transitions are unidirectional. Most derivations of fluctuation theorems are
based on the ability to map a realization onto a time-reversed counterpart, implicitly assuming that such a counterpart
exists. The absence of anti-resetting transitions therefore means that the derivations based on this assumption break down, and
that the usual fluctuation theorem will be typically violated in systems with resetting. In this paper we nevertheless
show that stochastic resetting systems satisfy two integral fluctuation theorems (IFTs). The first is the
Hatano-Sasa relation \cite{HatanoSasa2001,Trepagnier}, which was derived for dynamics without resetting.
It pertains to a processes
in which a system is driven from one steady-state to another by the modulation of parameters.
The derivation is illustrated
in \sref{sec:HSFT}. The thermodynamic interpretation of the Hatano-Sasa functional is discussed in \sref{sec:func}.
The second relation we derive is an extension of an IFT derived for Markov jump processes with
unidirectional transitions in \cite{Rahav2014}.
The fluctuating quantity appearing in this IFT is an interesting combination
of thermodynamic and dynamic quantities. The derivation of this second IFT is presented in \sref{sec:IFTDR}.
 We discuss
how the presence of resetting affects the
physical interpretation of the exponentially weighted functional that appears in each of the IFTs.
We also present numerical simulations to support the validity of both IFTs.
We discuss some implications of our results in \sref{Conclusion}.


\section{The Hatano-Sasa integral fluctuation theorem }
\label{sec:HSFT}

In 2001, Hatano and Sasa derived an insightful fluctuation relation for overdamped Langevin systems (without resetting) which are
driven from one steady-state to another
\cite{HatanoSasa2001}.
Consider a system that is prepared at a steady state matching an initial value of
some parameter $\alpha (0)$. The system is then driven by varying $\alpha$ with time, leading to a finite-time transition
between steady-states in which the actual time dependent probability distribution lags behind the distribution
of the momentary steady-state (with parameter $\alpha (t)$). On the other hand, for quasistatic variation
of the parameters the system moves through a continuous sequence of stationary
states since the lag between the actual distribution and the
momentary steady state vanishes. A quantitative measure of such a lag between two equilibrium states
is given by the Clausius inequality. The goal of Hatano and Sasa was to find an analogue in the case of steady states.

Hatano and Sasa identified a functional $Y[x(t)]$ of realizations of this process $x(t)$ that satisfies
an integral fluctuation theorem
\begin{equation}
\langle ~e^{-Y} ~\rangle~=~1,
\l{Hatano-Sasa-IFT}
\end{equation}
where angular brackets denote the average over an ensemble of realizations
of the process.
The functional is given by
\begin{equation}
Y[x(t);\alpha(t)]=\int_{0}^{\tau}~dt~\dot{\alpha}(t)~\frac{\partial \phi}{\partial \alpha}(x(t);\alpha(t)),
\l{Hatano-Sasa-functional}
\end{equation}
where $\phi (x;\alpha) \equiv - \ln \rho(x;\alpha)$
is the logarithm of the momentary steady-state distribution $\rho(x;\alpha)$ for a given value of $\alpha(t)$ \cite{HatanoSasa2001,Trepagnier}.

Hatano and Sasa proceeded to show that the functional $Y$ can be recast in a way that has an interesting thermodynamic
interpretation
\begin{equation}
Y=\beta Q^{\text{ex}}\left[ x(t) \right]+\Delta \phi\left[ x(t) \right],
\l{definition of Y}
\end{equation}
where $\beta^{-1}$ is the temperature of the ambient medium, and $\Delta \phi  = \phi(x(\tau);\alpha (\tau))-\phi(x(0);\alpha (0))$
is the difference between the final and initial values of $\phi$ along the realization.
$Q^{\text{ex}}$ was identified as the excess heat following Oono and Paniconi
who studied heat dissipation in non-equilibrium steady states in \cite{OonoPaniconi}.
This definition follows from a decomposition of
total heat into
excess heat, which is produced only during
transitions between steady-states, and a housekeeping heat, which is constantly produced to maintain
the steady state, so that $Q=Q^{\text{hk}}+Q^{\text{ex}}$. In equilibrium the housekeeping
heat vanishes and the excess heat
becomes identical to the total heat.
The Hatano-Sasa relation implies that
\begin{equation}
 \beta \langle Q^{\text{ex}} \rangle+ \langle \Delta \phi\ \rangle \geq 0 ,
 \l{second-law}
\end{equation}
which obtained from Eq.\eref{Hatano-Sasa-IFT} with the help of Jensen's inequality.
The excess heat is minimal for quasistatic processes where the external variation is slow
compared to other time-scales, and system
is effectively at the momentary steady state at each time step. For such processes $ \langle Q^{\text{ex}} \rangle=-T\langle \Delta \phi\ \rangle$.
Eq.\eref{second-law} can be recast as the second law for transitions between two steady states
if one notes that the Shannon entropy can be defined as $S(\alpha)=-\int~ dx ~\rho(x;\alpha) ~\ln \rho(x;\alpha)$,
leading to a Clausius-like inequality  $T \Delta S~\geq~-\langle Q^{\text{ex}} \rangle$ \cite{HatanoSasa2001}.

In this section we argue that Hatano-Sasa relation holds also
for dynamics with resetting. In fact, the derivation presented by Hatano and Sasa
in Ref. \cite{HatanoSasa2001} holds without changes. We present the derivation below
for completion and also to highlight a subtle point that needs some care when resetting is present.
Consider an overdamped particle diffusing in a parameter dependent potential landscape $U(x,\alpha)$.
The particle can also be reset to a fixed position $x_r$. This process occurs at a rate $r$ which is
taken to be spatially independent for simplicity. The probability distribution
of finding the particle in different locations $p(x,t;\vec{\lambda})$
evolves according to a Fokker-Planck equation
\begin{equation}
\label{eq:FPeq}
  \frac{\partial p}{\partial t}  = D \frac{\partial^2  p}{\partial x^2}
  +  \frac{\partial}{\partial x} \left[ \frac{\partial U}{\partial x} p \right]- r  p + r \delta (x-x_r).
\end{equation}
Here $\vec{\lambda} = \left\{ \alpha, r, x_r\right\}$ is the set of the process parameters that can be controlled externally,
and $D$ is the diffusion constant.
 If the parameters are not varied in time
the system will decay to a steady state with distribution
$\rho (x;\vec{\lambda})=e^{-\phi (x;\vec{\lambda})}$.

We now imagine that the system is initially in a steady state. At time $t \geq 0$, the system is driven out
of this state via variation of some of the parameters in time according to a known protocol, $\vec{\lambda} (t)$ (where $0\le t \le \tau$).
We further assume that the parameters are varied in a smooth manner. For the purpose of the derivation we divide the time interval
into many small time segments of size $\tau/N$. The original process can now be approximated by a process in which the
parameters are kept constant in each time step and are changed suddenly in between the time steps. Specifically, we take
$\vec{\lambda}(t)=\vec{\lambda}_k$ for $t_k < t < t_{k+1}$ with $0\leq k \leq N-1$ and $t_0 \equiv 0,~t_N\equiv \tau$.
For larger and larger  values of $N$ this piecewise constant process will be a better and better approximation of the original process.
Let us denote the transition probability between two states in one unit of time $\tau/N$
for a fixed $\vec{\lambda}$ by $P(x^\prime | x ; \vec{\lambda})$.
By definition this propagator maps the steady state distribution onto itself
 $ \rho (x^\prime;\vec{\lambda}) = \int dx ~P(x^\prime | x ; \vec{\lambda}) \rho (x;\vec{\lambda}).$

One can now define functionals $G [x(t);\vec{\lambda} (t) ]$ over realizations using a limiting procedure where
$G$ is expressed in terms of the values of $x(t)$ at each time step $t_k=k/N \tau$ and then by taking the limit $N \rightarrow \infty$.
The ensemble average of this functional for a given value of $N$ is given by
\bea
\label{eq;deffunc}
\langle G \rangle \simeq \int~ \prod_{k=0}^{N} dx_k~\bigg( \prod_{k=0}^{N-1}~P(x_{k+1}|x_k;\vec{\lambda}_k) \bigg)
\rho(x_0;\vec{\lambda}_0) G[x(t);\vec{\lambda}(t)].
\eea
When the limit $N \rightarrow \infty$ is taken (with fixed $\tau$) the piecewise constant process approaches
the original smooth process, while the functional converges to a limiting form.
The Hatano-Sasa fluctuation theorem is obtained by noting that the functional
\bea
\label{eq:defr}
{\cal R}\left[ x_k; \vec{\lambda}_k  \right] \equiv \prod_{k=0}^{N-1}~\f{\rho(x_{k+1};\vec{\lambda}_{k+1})}{\rho (x_{k+1};\vec{\lambda}_k)},
\eea
satisfies a relation
\bea
\Big\langle ~{\cal R}\left[ x_k; \vec{\lambda}_k  \right]~\Big\rangle \simeq 1.
\l{eq:relationr}
\eea
This can be verified by direct substitution of Eq. (\ref{eq:defr}) into Eq. (\ref{eq;deffunc}).
Writing ${\cal R}$ as an exponent of a functional, using the definition of $\phi$, and taking the limit $N \rightarrow \infty$ results
in
\begin{equation}\label{eq:HSwithr}
  \left< \exp \left[ - \int_{0}^{\tau} dt ~\dot{\vec{\lambda}} \cdot \frac{\partial \phi (x;\vec{\lambda})}{\partial \vec{\lambda}}\right]\right>=1.
\end{equation}
Here $ \dot{\vec{\lambda}} \cdot \frac{\partial}{\partial \vec{\lambda}}  \equiv \dot{\alpha} \frac{\partial}{\partial \alpha }
+  \dot{r} \frac{\partial}{\partial r }+ \dot{x}_r\frac{\partial}{\partial x_r }$.
The Hatano and Sasa functional as appeared in \cite{HatanoSasa2001}  is recovered
when only $\alpha$ is varied, but one can see the resulting fluctuation theorem
also holds when the parameters characterizing the resetting process are varied.
One interesting difference between dynamics with and without resetting is that realizations with resetting
need not be continuous. In fact, the resetting events involve finite and
sudden changes in the particle position. Nevertheless, the dependence of $\phi (x;\vec{\lambda})$ on
parameters is smooth, and as a result the derivatives appearing in Eq. (\ref{eq:HSwithr}) are well defined.
In contrast, recasting the functional in a form that would involve derivative with respect to $x$ must
be done with proper care. This will become
important when we try to give a physical interpretation to the functional appearing in (\ref{eq:HSwithr}). This is the subject of the next section.


\begin{figure}
\centering
\begin{subfigure}{0.56\textwidth}
  \centering
  \includegraphics[width=\linewidth]{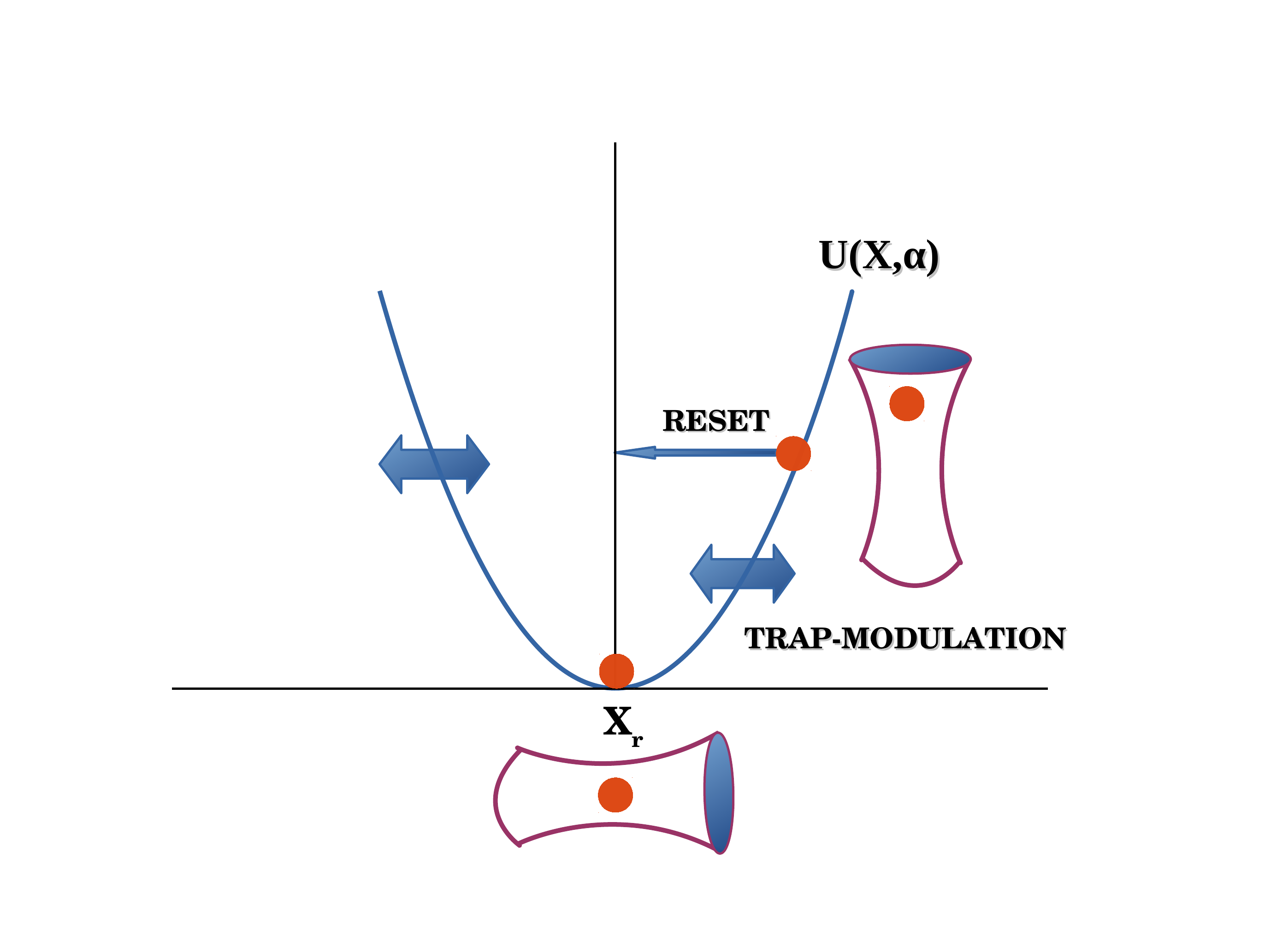}
\end{subfigure}
\begin{subfigure}{.42\textwidth}
  \centering
  \includegraphics[width=\linewidth]{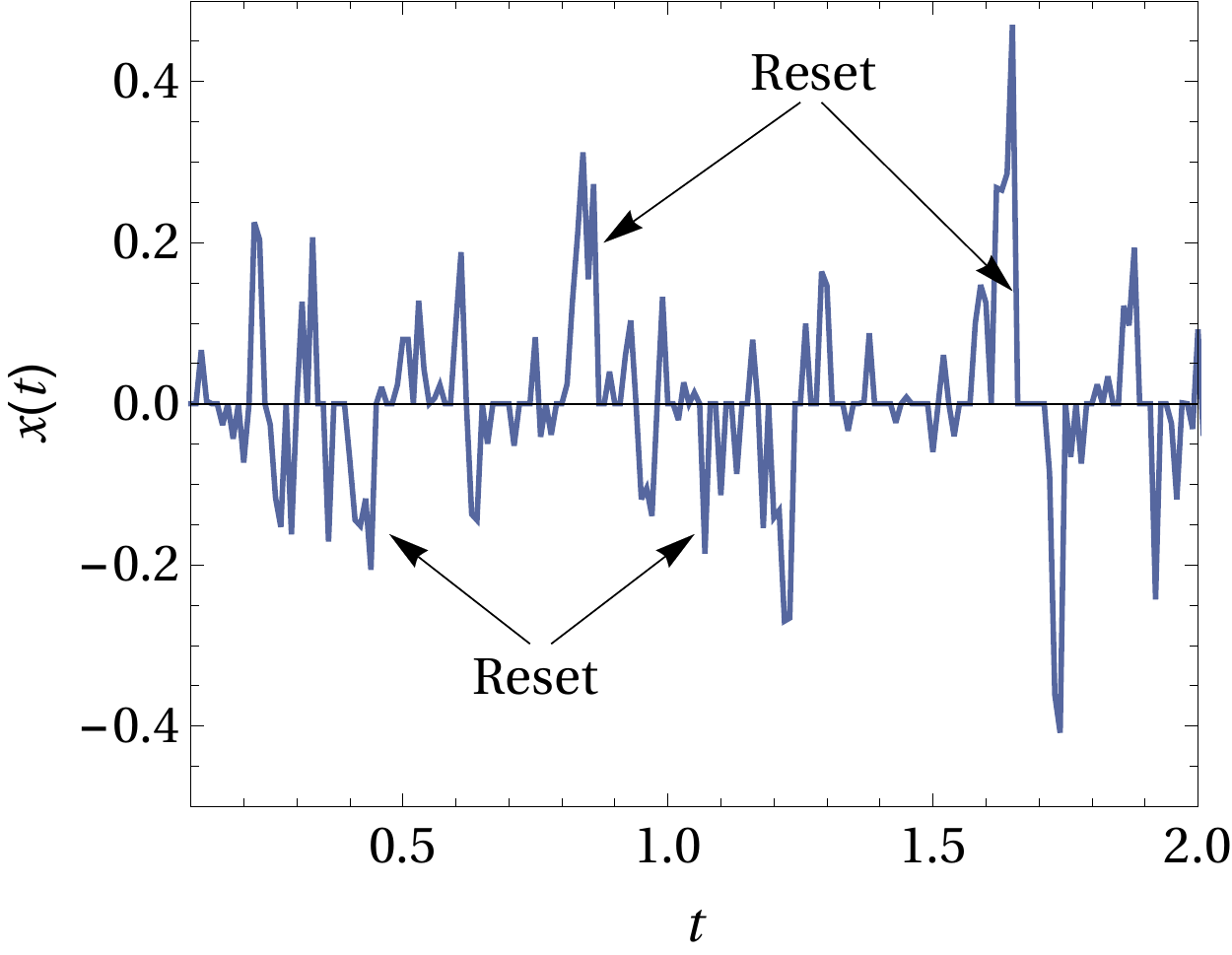}
\end{subfigure}%
\caption{ \textbf{Left panel}: Schematic of an elementary Brownian motion in a potential $U(x,\alpha)$ subjected to a resetting mechanism.
The potential is varied by external modulation.
\textbf{Right panel}: A single trajectory of such a Brownian particle
with resetting jumps to the origin is shown. The reset events are indicated within the panel. }
\l{fig:test}
\end{figure}


\section{Thermodynamic interpretation of the Hatano-Sasa functional in the presence of resetting}
\label{sec:func}

The physical interpretation of the functional $Y[ x(t);\vec{\lambda}(t)]$ is not obvious at first sight.
For diffusive dynamics without resetting Hatano and Sasa showed that this functional expresses the excess heat,
namely heat exchanged between the diffusing particle and its environment beyond the heat that would have been
exchanged had the system been maintained at steady state.
The ensemble average of the latter is the heat required to
keep the system at a steady state and is therefore termed the house-keeping heat. The functional $Y[ x(t);\vec{\lambda}(t) ]$
is therefore a measure for the deviation from quasistatic time-variation of parameters. In this section we shall see
that this qualitative picture applies also to systems with resetting, but with some important differences.

For this purpose we examine a single realization of a process where the particle diffuses in a potential
$U(x,\alpha)$ but gets interrupted with resetting to $x_r$. The microscopic dynamics at an infinitesimal time step $dt$ is given by
\begin{equation}\label{eq:defdx}
  dx = x(t+dt)-x(t)= \hat{N}_t \left[ x_r-x(t)\right]+\left( 1-\hat{N}_t\right) \left[ -
  \frac{\partial U}{\partial x}~ dt+\sqrt{2D}~ d B_t\right].
\end{equation}
Here $x(t)$ is the position of the particle, while $\hat{N}_t$ is a random variable which determines whether there was a resetting event in
the time interval
between $t$ and $t+dt$. It can take only two values, $0$ or $1$. The probability for a reset event is given by
$P(\hat{N}_t = 1 )=r dt$, whereas the complementary probability is $P(\hat{N}_t = 0)=1-rdt$. The change in particle position due to diffusion is simply
$d x_{\text{diff}} \equiv - \frac{\partial U}{\partial x}~ dt+\sqrt{2D} ~d B_t$,
where $dB_t$ is a Wiener process. $dB_t$ therefore satisfies
 $\left< dB_t \right>=0$, $\left< dB_t^2\right> = dt$, and $\left< dB_t dB_{t^\prime} \right> =0$
for $t \ne t^\prime$ (assuming non overlapping time intervals).
The stochastic dynamics of systems with resetting exhibits an interesting feature which is absent in purely diffusive dynamics.
The difference $dx$ needs not be small in an infinitesimal time step. The reason for this is obvious. In reset events
the particle position is changed suddenly so that $dx=x_r-x(t)$ can be arbitrarily large. As a result one should take care when making
manipulations that require expansions in powers of $dx$.

In their paper Hatano and Sasa used integration by parts to rewrite the functional in a more physically transparent form. This is precisely the type of
manipulation that can be problematic at reset events. However, we note that the probability of a reset at each time step is infinitesimal.
The random variables $\hat{N}_t$ that determine the epochs of reset constitute a Bernoulli process which furthermore converges to
a Poisson process when $dt \rightarrow 0$. As a result, the probability to find $J_r$ resetting events in a realization of duration $\tau$
is
\begin{equation}\label{eq:poisson1}
  P(J_r,\tau) = \frac{(r\tau)^{J_r}}{J_r!} e^{-r \tau},
\end{equation}
while the waiting time between two consecutive reset events is distributed according to
$P(\Delta t) =r e^{-r\Delta t}$.
One sees that averages over realizations are dominated by realizations with a finite number of separate resetting events.
The relative weight of realizations with an extremely large number of resetting events is
therefore negligible. One can then focus on realizations with a finite number of separate resetting events.

We thus examine the functional $Y[x(t);\vec{\lambda}(t)]$ in Eq. (\ref{eq:HSwithr}) for a realization $x(t)$ that has a finite number $J_r$
of resettings at times $0<t_{\tiny{1}}<t_2<\cdots<t_{J_r}<\tau$. The integrand in Eq. (\ref{eq:HSwithr}) may change suddenly in the vicinity
of the resetting points,
but the change is a jump between two finite values.
As a result excluding a finite number of infinitesimal time segments around the resetting times will not change
the value of the functional. One can therefore rewrite the functional as
\begin{equation}\label{eq:func2}
  Y[x(t);\vec{\lambda}(t)] = \int_{0}^{t_1^-} dt \dot{\vec{\lambda}} \cdot \frac{\partial \phi}{\partial \vec{\lambda}} +
  \int_{t_1^+}^{t_2^-} dt \dot{\vec{\lambda}} \cdot \frac{\partial \phi}{\partial \vec{\lambda}} + \cdots +
  \int_{t_{n_r}^+}^{\tau} dt \dot{\vec{\lambda}} \cdot \frac{\partial \phi}{\partial \vec{\lambda}}~,
\end{equation}
where $t_i^{-},~t_i^{+}$ denote the times just before and after the resetting events. Since resetting is instantaneous,
the difference between these two times is infinitesimal.
In each of the time segments in Eq. (\ref{eq:func2}) the particle performs diffusion without resetting. As a result, we
can use integration by parts to obtain
\begin{multline}\label{eq:func3}
   Y[x(t);\vec{\lambda}(t)] = \phi(x(t_1^-); \vec{\lambda} (t_1))-\phi (x_r;\vec{\lambda} (0))
   - \int_{0}^{t_1^-} dt \frac{dx}{dt} \frac{\partial \phi}{\partial x} + \phi(x(t_2^-); \vec{\lambda} (t_2))
   -\phi (x_r;\vec{\lambda} (t_1)) - \int_{t_1^+}^{t_2^-} dt \frac{dx}{dt} \frac{\partial \phi}{\partial x} \\
   \cdots + \phi(x(\tau); \vec{\lambda} (\tau))-\phi (x_r;\vec{\lambda} (t_{J_r})) - \int_{t_{J_r}^+}^{\tau} dt \frac{dx}{dt}
   \frac{\partial \phi}{\partial x}~,
\end{multline}
Our use of integration by parts means that the stochastic integrals in Eq. (\ref{eq:func3}) should be interpreted according
to the Stratonovich prescription \cite{Gardiner}.

We are now in position to recast the functional $Y$ in terms of thermodynamic quantities. The thermodynamic interpretation
of stochastic resetting was discussed in by Fuchs {\it et al.} \cite{Seifert2016}. For instance, they noted that
the resetting step involves a resetting work of $U(x_r,\alpha)-U(x(t^-),\alpha)$ that is done on the system. In addition,
each resetting step must also be associated with a change of the fluctuating entropy of the system, $-\ln p(x,t;\vec{\lambda})$
where $p$ satisfies Eq.\eref{eq:FPeq}. In a seminal paper Seifert has shown that
inclusion of such fluctuating system's entropy in the total entropy production results in exact, rather than asymptotic
fluctuation theorems \cite{Seifert2005}.
The change of this fluctuating entropy in a single
resetting step (suppressing the explicit time dependence in $p$) is
\begin{equation}\label{eq:defflent}
  \Delta S_{\text{reset}}[x(t);\vec{\lambda}(t)] = \ln \frac{p(x(t^-);\vec{\lambda}(t))}{p(x_r;\vec{\lambda} (t))}.
\end{equation}
Fuchs {\it et al.} discussed the ensemble average of this quantity and showed that this contribution for the entropy production
must enter the second law of thermodynamics of resetting systems, see Eqs. (10)-(13) in Ref. \cite{Seifert2016}.

Examination of the functional (\ref{eq:func3}) shows that it has a sum of contribution of the form
$\phi (x(t_i^-);\vec{\lambda}(t_i))-\phi (x_r;\vec{\lambda}(t_i))$ from
all the reset events (as shown in \fref{fig:test} for instance)  along the realization.
This can be identified as a change in entropy due to resetting,
but with the momentary steady state distribution $\rho$ replacing
the actual probability distribution $p$
\begin{equation}
\label{eq:defexcessentropy}
  \Delta S^{\text{ex}}_{\text{reset}} [x(t);\vec{\lambda}(t)]  \equiv   \sum_{i=1}^{J_r} \Big[\phi(x_r;\vec{\lambda}(t_i))
  - \phi(x (t_i^-);\vec{\lambda}(t_i)) \Big]=
  \sum_{i=1}^{J_r}~\ln~\frac{\rho(x (t_i^-);\vec{\lambda}(t_i))}{\rho(x_r;\vec{\lambda}(t_i))},
\end{equation}
where we have used the definition of $\phi$.
 The summation runs over all resetting events in the realization $[x(t);~0\leq t \leq \tau$].
 We use the notation {\em ex} for this measure of entropy production to conform with
 customary notation (See e.g. \cite{HatanoSasa2001,OonoPaniconi,Speck2005}).
 As will be discussed later, for systems with resetting, the interpretation
 of $\Delta S^{\text{ex}}_{\text{reset}}$ as an excess quantity is somewhat misleading.
We note that the mean rate of this resetting entropy production is given by
\bea
\dot{S}^{\text{ex}}_{\text{reset}}=r\int dx~p(x;\vec{\lambda})~
\ln \frac{\rho(x ;\vec{\lambda})}{\rho(x_r;\vec{\lambda})}.
\label{eq:defexcessentropyPR}
\eea
The functional $Y$ can now be rewritten as
\begin{equation}\label{eq:func4}
    Y [x(t);\vec{\lambda}(t)] = \Delta \phi -  \Delta S^{\text{ex}}_{\text{reset}} - 
    \int_{0}^{t_1^-} dt \frac{dx}{dt} \frac{\partial \phi}{\partial x}
    - \int_{t_1^+}^{t_2^-} dt \frac{dx}{dt} \frac{\partial \phi}{\partial x} \cdots -
    \int_{t_{J_r}^+}^{\tau} dt \frac{dx}{dt} \frac{\partial \phi}{\partial x}.
\end{equation}

The stochastic integrals in Eq. (\ref{eq:func4}) can be expressed in terms of the excess heat produced during a realization in a similar way to the
approach taken by Hatano and Sasa \cite{HatanoSasa2001}. Heat is exchanged between the system and the thermal reservoir
only when the particle diffuses and this is given by
\begin{equation}
  Q [x(t);\vec{\lambda}(t)] = -\int_{0}^{t_1^-} dt \frac{dx}{dt} \frac{\partial U}{\partial x} -\int_{t_1^+}^{t_2^-} dt
  \frac{dx}{dt} \frac{\partial U}{\partial x} \cdots -\int_{t_{J_r}^+}^{\tau} dt \frac{dx}{dt} \frac{\partial U}{\partial x},
  \l{total-heat}
\end{equation}
which is based on the fact that in overdamped systems the force $-\frac{\partial U}{\partial x}$ has to balance the force that the
particle in the environment apply on the diffusing particle. The stochastic integrals here
should also be interpreted according to the Stratonovich prescription.
(See e.g. Ref. \cite{Sekimoto-book} for a more detailed discussion on the differences between the Ito and Stratonovich prescriptions in the
context of stochastic thermodynamics).
It will be now useful to define the housekeeping heat along the trajectory in the following way
\cite{HatanoSasa2001,OonoPaniconi,Speck2005}
\begin{equation}\label{eq:defQhk}
  Q^{\text{hk}}  [x(t);\vec{\lambda}(t)] =  \int dt~ v_{\text{ss}} (x(t); \vec{\lambda})~ \frac{dx}{dt},
\end{equation}
where $ v_{\text{ss}} (x; \vec{\lambda})= J_{\text{ss}} (x;\vec{\lambda})/\rho (x;\vec{\lambda})$ is the mean local velocity
of particles at the steady state distribution with the external parameters $\vec{\lambda}$.
The diffusive particle current is given by $J_{\text{ss}}(x;\vec{\lambda}) = - \frac{\partial U}{\partial x} \rho - D \frac{\partial \rho}{\partial x}$
also for diffusive dynamics with resetting. The main difference between the current case and that of systems without resetting is that here the current
$J_{\text{ss}}$
will generally be position dependent. Substitution of the expression for the current allows to express the house keeping heat as
\begin{equation}\label{eq:ekq2}
   Q^{\text{hk}}  [x(t);\vec{\lambda}(t)] = \int dt \left(-\frac{\partial U}{\partial x} + \frac{1}{\beta} \frac{\partial \phi}{\partial x}\right)
   \frac{dx}{dt}.
\end{equation}
The stochastic integrals in Eq. (\ref{eq:func4}) are clearly the difference between
the heat of a realization and its house keeping counterpart from Eq.\eref{eq:ekq2}, namely,
\begin{equation}\label{eq:finalfunc}
  Y [x(t);\vec{\lambda}(t)]= \Delta \phi - \Delta S^{\text{ex}}_{\text{reset}} + \beta Q^{\text{ex}},
\end{equation}
where the excess heat is defined as the difference $Q^{\text{ex}} \equiv Q-Q^{\text{hk}}$.

In absence of resetting the so-called excess heat behaves like a proper excess quantity. By this
we mean that its ensemble average is approximately proportional to $\Delta \phi$ in slow, gradual, processes. It
does not grow with the duration of the process. This is no longer true for the excess heat in the presence of resetting.
This can be understood intuitively. The resetting dynamics is built out of a sequence of resetting
events and periods of diffusion. The systematic bias of the former,
due to the fact that resetting events always put the particle as $x_r$, indicates that the diffusion
will also have a preferred direction. Both the mean
excess heat and the mean resetting entropy are therefore expected to grow with time even for systems at steady state. Nevertheless, the derivation
above shows that the combination $\beta Q^{\text{ex}} -  \Delta S^{\text{ex}}_{\text{reset}} $ is the one
which behaves like a proper excess quantity, namely that it has a mean that is not proportional to the 
duration in quasistatic processes.
One should nevertheless note that separating this well defined excess quantity into resetting and heat related parts results in 
partial contributions which are expected to behave awkwardly.
We finally obtain a second law-like inequality using the Jensen's relation in Eq.\eref{eq:HSwithr},
\bea
\beta \langle Q^{\text{ex}} \rangle+ \langle \Delta \phi\ \rangle - \langle  \Delta S^{\text{ex}}_{\text{reset}}  \rangle \geq 0,
\eea
for transitions between steady states in systems with resetting.

To summarize,
the derivation presented in Secs. \ref{sec:HSFT} and \ref{sec:func} show that the Hatano-Sasa
integral fluctuation theorem is also valid for systems with resetting. Its derivation is almost unchanged by the inclusion of resetting.
A bit of care is needed since the
functional $Y [x(t);\vec{\lambda}(t)]$ in Eq.\eref{eq:HSwithr}
includes realizations with intermittent long range jumps due to resetting.

\begin{figure}[t]
\includegraphics[width=.3\hsize]{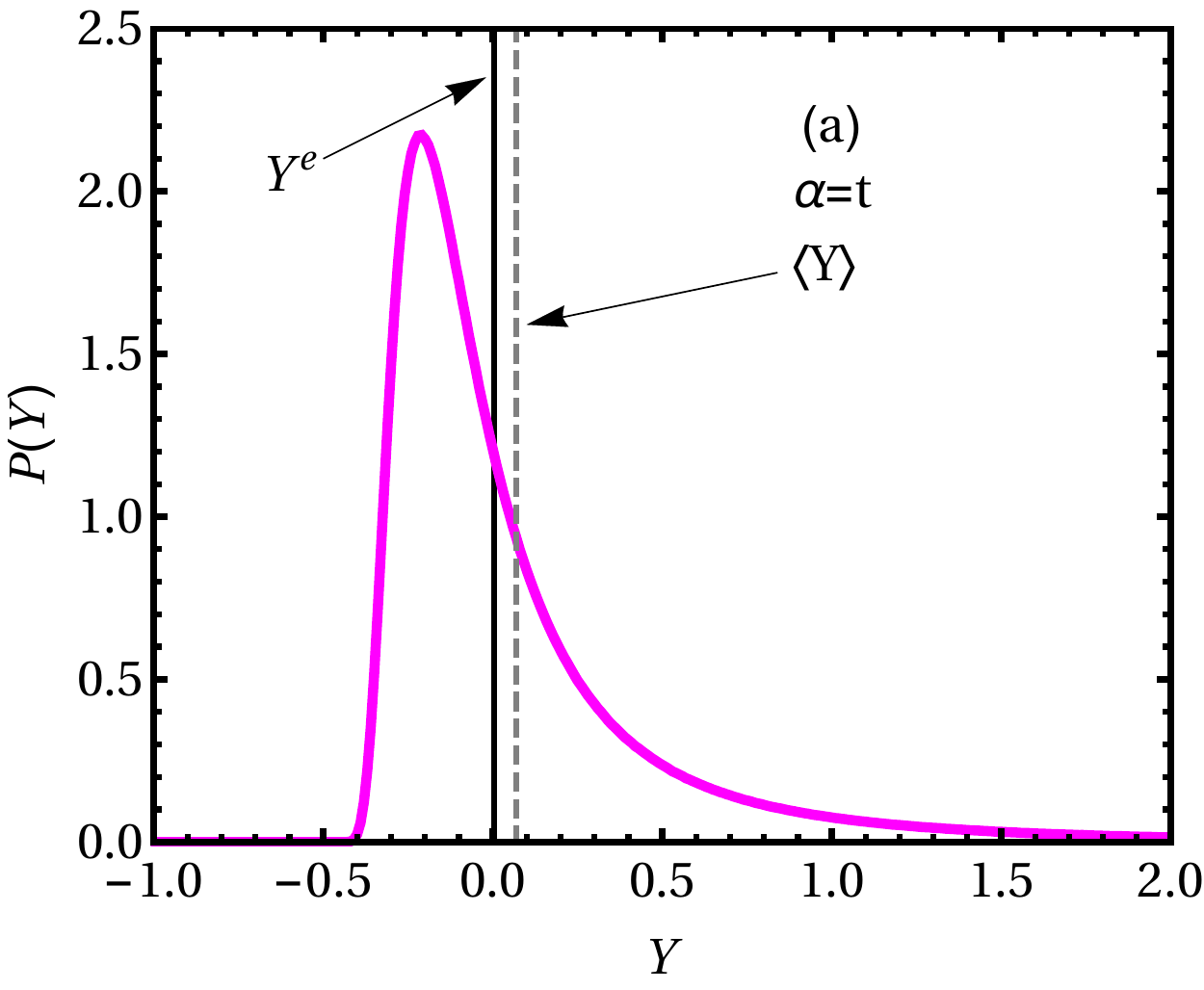}
\includegraphics[width=.3\hsize]{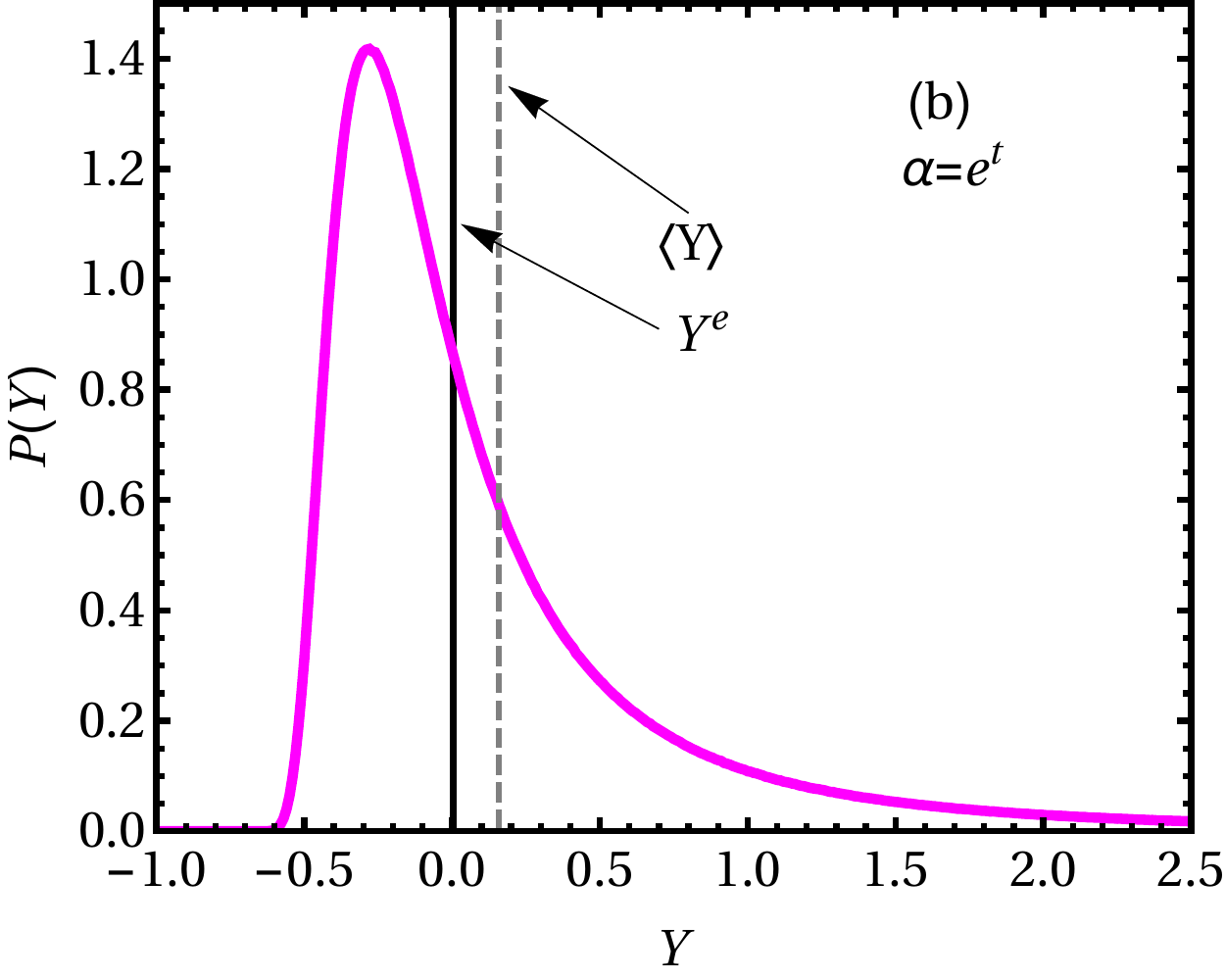}
\includegraphics[width=.29\hsize]{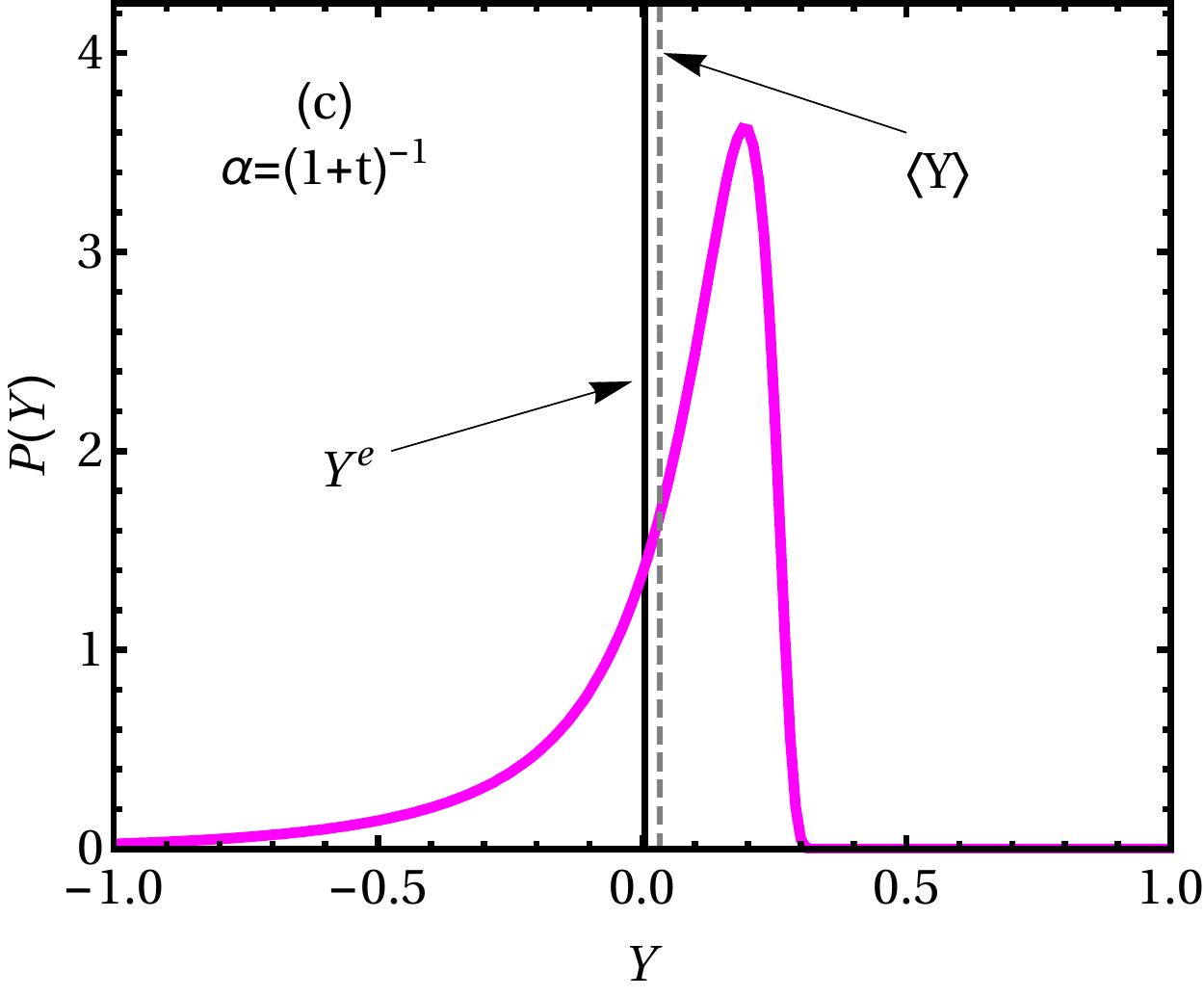}
\caption{Numerical distribution of the Hatano Sasa functional $Y$ of an overdamped Brownian particle diffusing in a potential
$U(x,\alpha)=\alpha(t)|x|$ in the presence of stochastic resetting.
The system is prepared in a steady state for $\alpha=1.0$ and $r=0.6$ at $t=0$. The diffusion constant is fixed at $D=1/2$, with $D\beta=1$.
We drive the system out of this steady state
by varying $\alpha$  with time. We employed three different protocols (a) $ \alpha(t)=t$, (b) $\alpha(t)=e^{t}$, and (c) $ \alpha(t)=(1+t)^{-1}$
respectively, where
$0 \le t \le 1$.
For each realization of the process we computed a value of the Hatano-Sasa functional using Eq. (\ref{HSfunctionaldiscretize}).
The distributions depicted in the figure were generated from $N_r=10^8$ realizations of the process.
 Both positive and negative values of $Y$ are observed,
but we find that $\langle Y \rangle$ is always positive, as expected from the Hatano-Sasa FT. This is shown by the dashed vertical line (in grey)
in each panel. The solid vertical lines (in black) correspond to the value of
$Y^e=\ln \frac{1}{N_r} \sum_{i=1}^{N_r}e^{-Y_i}$, where the sum runs over all realizations. These lines are located at the coordinates
$0.004, 0.002$ and $0.005$ in panels (a), (b), and (c) respectively. }
\l{fig1}
\end{figure}

\subsection{Numerical Simulations}
To illustrate our results we performed simulations of a
simple example of stochastic dynamics with resetting.
Specifically,
we considered an overdamped particle
diffusing in a potential $U(x,\alpha)=\alpha(t) |x|$, where $\alpha(t)$ is the stiffness
of the potential.
The whole system is immersed in a bath with temperature $\beta^{-1}$.
At each microscopic time step $dt$, the system may be reset to the origin
with probability $rdt$. Alternatively, the system diffuses for a time $dt$ with the complementary probability $1-rdt$. 
The diffusion constant has been fixed at $D=1/2$ with $D\beta=1$.
The system is initially prepared at a steady-state with $\alpha=1.0, r=0.6$. It is then driven away
from this steady state using a time variation of $\alpha$, where $0 \le t \le 1$.
We used three different driving protocols:
(a) $ \alpha(t)=t$, (b) $\alpha(t)=e^{t}$ and (c) $ \alpha(t)=(1+t)^{-1}$ respectively.
As $\alpha$ is varied externally, the system probability distribution follows a series of non-equilibrium states which
lag behind the momentary steady state.

Each realization of the stochastic process Eq.\eref{eq:defdx} is a fluctuating trajectory. For each such realization we calculate the
value of the functional $Y[x(t)]$ with the help of Eq.\eref{Hatano-Sasa-functional}.
This requires discretization of the integral appearing in the definition of $Y$, which is done using
\bea
Y=\sum_{i=0}^{N-1} \bigg\{ \phi(x_{i+1};\alpha_{i+1})-\phi(x_{i+1};\alpha_{i}) \bigg\}.
\l{HSfunctionaldiscretize}
\eea
 Evaluation of this functional requires knowledge of the steady state distribution
$\rho(x;\alpha)$ for any value of system and resetting parameters. For this particular system
analytical expressions for the steady state distribution are known \cite{Pal14}, and they have been
employed in our calculations.

Fig. \ref{fig1} presents the resulting distribution function of the values of $Y$
for the three processes mentioned above. They are generated from
histograms of $N_r=10^8$ realizations of the dynamics. The Hatano-Sasa relation is verified
by computing $Y^e \equiv \ln \frac{1}{N_r}\sum_{i=1}^{N_r} e^{-Y_i} $ for each one the different driving protocols.
The solid vertical lines in Fig. \ref{fig1} depict the value of $Y^e$ whereas the vertical dashed lines correspond to the value of $\langle Y \rangle$.
Our numerical simulation returns values of $Y_e \simeq 0.002 - 0.005$ which is consistent with the
predictions of the Hatano-Sasa relation, $Y_e = 0$.


\section{Integral FT for discrete jump processes with resetting}
\label{sec:IFTDR}

In this section, we present another integral fluctuation theorem
which holds for Markov jump processes.
In this setup resetting is introduced by including unidirectional transitions
that point towards a specific resetting site.
The fluctuation theorem commonly emerges from a comparison of the
probabilities of a realization and that of its time reversed counterpart.
When the dynamics exhibits microreversibility each allowed trajectory has a time-reversed counterpart
and vice-versa. Importantly, the mapping between trajectories and their time-reversed counterparts is one-to-one.
Dynamics with resetting
violates microreversibility. All resetting transitions
put the particle at the reset site, whereas the dynamics does not include any \textit{anti-resetting} transitions.
Despite this, we show that by introducing an auxiliary dynamics with
inverted (anti-resetting) transitions a one-to-one mapping of realizations
can still be achieved.
This allows us to derive an IFT
which is solely given in terms of the original resetting dynamics.

\begin{figure}[t]
\centering
\begin{subfigure}{.560\textwidth}
  \centering
  \includegraphics[width=1.1\linewidth]{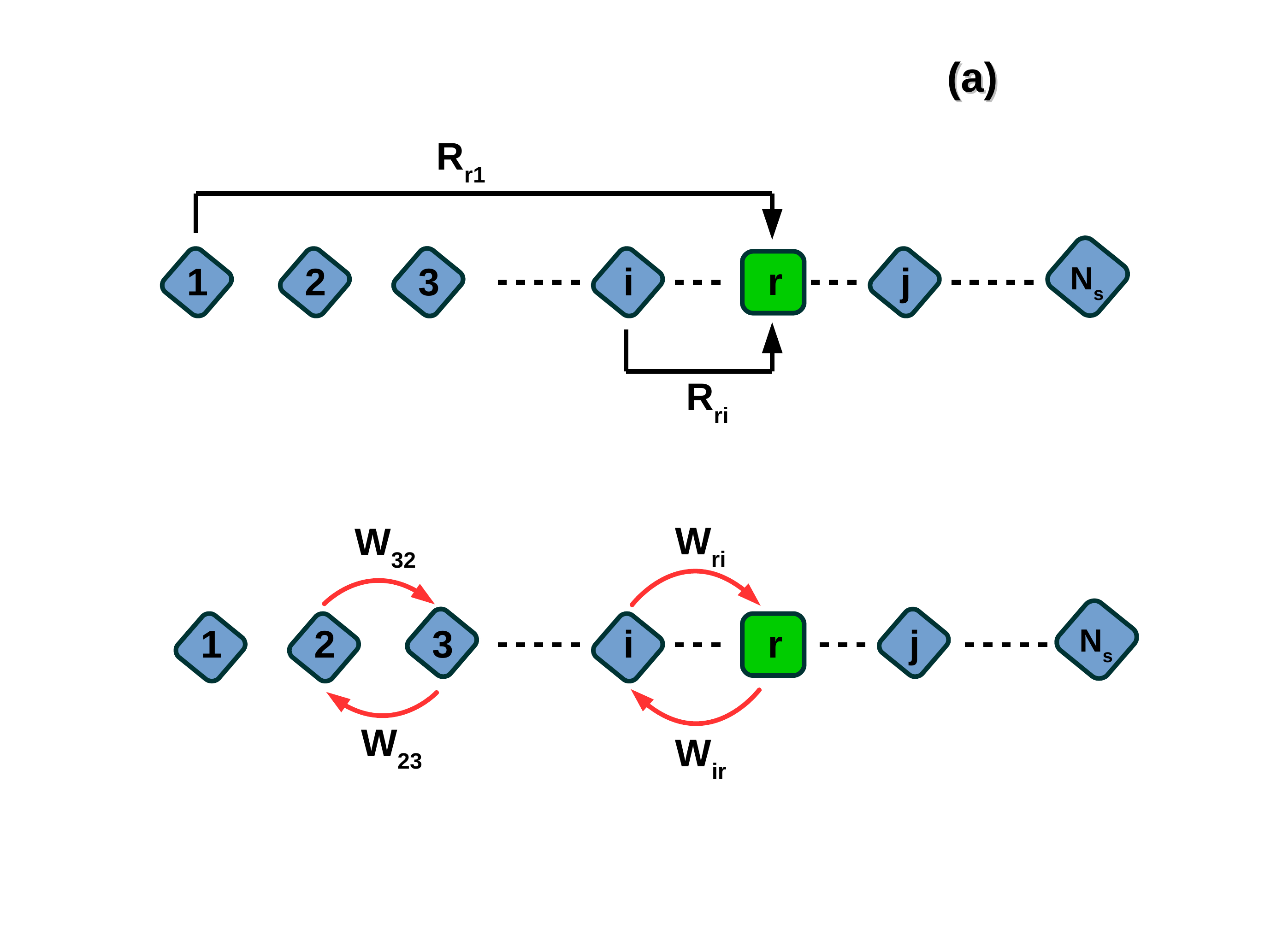}
\end{subfigure}%
\begin{subfigure}{0.576\textwidth}
  \centering
  \includegraphics[width=1.0\linewidth]{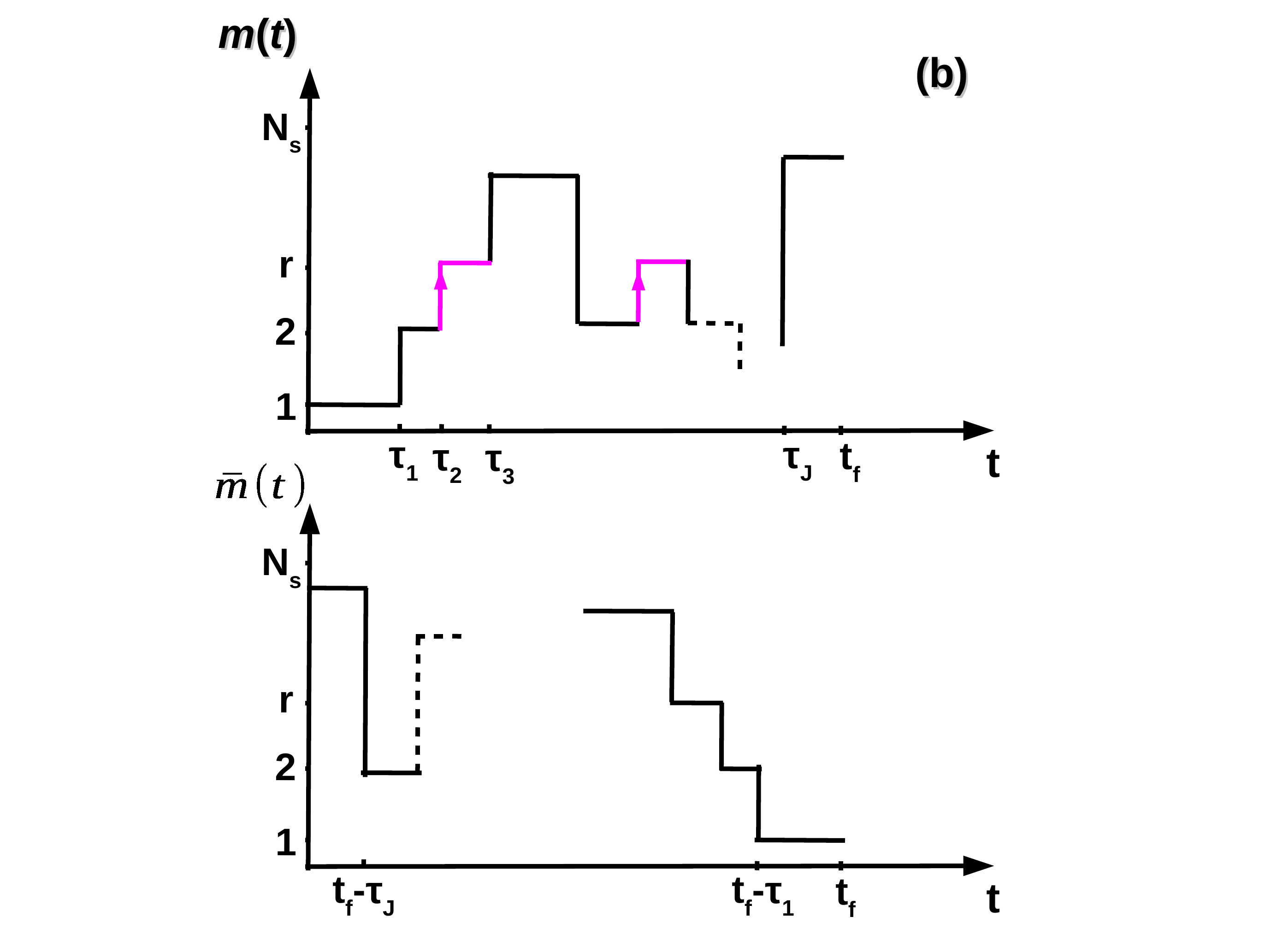}
\end{subfigure}
\caption{\textbf{Left panel (a)}: Schematic depiction of the discrete jump process with resetting. The resetting site $r$ is denoted by an orange square
while the other sites are depicted as blue squares. Between any two sites $(i,j)$ in the system, there are bidirectional transitions
$(W_{ij}, W_{ji})$ indicated by red arrows. In addition, there are unidirectional resetting transitions from all the sites to the resetting
site denoted by $R_{ri}$. Both type of transitions exist simultaneously. They are drawn separately to avoid cluttering.
\textbf{Right panel (b)}: Heuristic representation of a single realizations of the jump process shown on the left panel.
$m(t)$ corresponds to the site that the system
occupies at time $t$. A time reversed trajectory $\overline{m}(t)$ as defined in the main text
has also been plotted in conjugation with the original trajectory.}
\label{figr3}
\end{figure}

We model the resetting dynamics as a jump process on a discrete lattice with the sites $\{1, 2, 3, ... r ..., N_s   \}$,
where the reset site is labeled by $r$. We distinguish between two physically distinct types of Markov transitions. The first type consists of
diffusive jumps between any two sites which occur
due to the interaction between the system and the ambient medium, possibly including some external bias. These
jumps are bidirectional.
The second type of transitions is the resetting events. These are transitions from any site (except the reset site)
to the particular
reset site.
We view the resetting events as done by some external agent.
These transitions are unidirectional, as there are no anti-resetting transitions.
To highlight this distinction we denote the rate of bidirectional jump from site $m$ to $n$ by $W_{nm}(t)$, and
the rate of
resetting transitions from site $m$ to $r$ by $R_{rm}(t)$
respectively.
The two types of transitions are schematically depicted in Fig. \ref{figr3}a.
The bi-directionality of the diffusive transitions is expressed by demanding that
$W_{nm}(t) >0$ implies that also $W_{mn} (t) >0$.
The probability to find the system in site $n$ at time $t$, denoted by $p(n,t)$,
evolves according to a master equation
$\f{dp}{dt}=\mathcal{L}p$, where  $\mathcal{L}$ is the
transition rate matrix.
The off-diagonal elements of the transition rate matrix are composed from the transition rates. Its diagonal
elements are chosen to ensure conservation of probability $\mathcal{L}_{ii} = - \sum_{j \ne i} \mathcal{L}_{ji}$.
The transition rates may be time-dependent.

Let us consider a particular realization of the jump process, $\Gamma= \left\{ m(t)\right\}$, evolving between $t=0$ and $t=t_f$,
where the state of the system $m(t)$
transitions between a sequence of states $m_j$, such that
$m(t) \equiv m_{j},$ for $\tau_j \leqslant t \leqslant \tau_{j+1}$.
In this notation
$m_0$ is the initial state of this particular realization, while the system
is at $m_J$ at the final time $t_f$.
This realization is heuristically depicted in Fig. \ref{figr3}b.
The Markovian nature of this jump process allows the construction
of the probability density of a realization from a few simple building blocks.

During the realization $\Gamma$ the system spends a finite amount of time
in a sequence of sites.
The probability that the system is at $m(t)=n$ for the time segment $(t_1,t_2)$, without making any transitions,
is the so-called survival probability. It is given by
\bea
S_{n}(t_2,t_1)= \exp \left[-\int_{t_1}^{t_2}  d\tau ~e_{n}(\tau) \right],
\l{survival-all-site}
\eea
where $e_n(t)$ is the total rate of transitions out of site $n$
\bea
e_{n}(t)=K_n(t)+ R_{rn}(t)\Big[ 1-\delta_{rn} \Big].
\l{exit-rate-all-site}
\eea
Here $K_n(t)=\sum_{m \neq n} W_{mn}(t)$ is the contribution of bidirectional transitions to this escape rate.
The time that a realization spends in site $n$ without leaving is therefore distributed according to
\bea
f_n(t)=e_n(t)\exp \left[-\int_{0}^{t}  dt ~e_{n}(\tau) \right].
\eea
The expression for the escape rate from the reset site $r$ is somewhat different, since there
are no resetting transitions out of this site.
It is
given by $e_{r}(t)=\sum_{m \neq r} W_{mr}(t)$.

Since the jump process is Markovian the probability density of the realization $\Gamma$ is given by
\bea
P\left[ \Gamma \right]=p_i(m_0)~S_{m_0}(\tau_1,0)W_{m_1m_0}(\tau_1)~S_{m_1}(\tau_2,\tau_1)R_{rm_1}(\tau_2)~
S_r(\tau_3,\tau_2)....W_{m_Jm_{J-1}}(\tau_J)~S_{m_J}(t_f,\tau_J).
\eea
where the initial condition is chosen randomly from a distribution $p_i$. This specific realization
includes a resetting event at $t=\tau_2$, as well as several diffusive transitions at times $\tau_1, \tau_3, \tau_4, \cdots$.
This probability density can be recast in a way that separates the roles of diffusing and resetting transitions
\bea
P\left[ \Gamma \right]=p_i(m_0) \prod_{j=1}^{J}
\exp\Big[-\int_{\tau_j}^{\tau_{j+1}} dt~ K_{m_j}(t)-(1-\delta_{rm_j})\int_{\tau_j}^{\tau_{j+1}} dt ~R_{rm_j}(t) \Big]
\prod_{j \in J_d}~W_{m_{j+1}m_j}(\tau_j)~\prod_{j \in J_r}~R_{rm_j}(\tau_j),
\l{forward-trajectory}
\eea
where $J_d$ ($J_r$) is the set of $j$ values of diffusive (resetting) jumps that occurred during the realizations.
Note that the second term in the exponential of Eq.\eref{forward-trajectory} only
picks up contributions when the actual state $m(t)$ is not at the reset site $r$.

To proceed we examine an auxiliary dynamics in which the resetting transitions are replaced by
\textit{anti-resetting transitions}. This means that the `resetting' site has many outgoing
\textit{anti-resetting transitions} with a total
rate of
$R^{\text{aux}}_r(t)=\sum_{m \neq r} \overline{R}_{mr}(t)$, where $\overline{R}_{mr}(t)$ is
the \textit{anti-resetting} rate from the site $r$ to site $m$.
For each realization $\Gamma$ of the original dynamics let us examine
a realization $\overline{\Gamma} \equiv \left\{ \overline{m}(t)\right\}=\left\{ m(t_f-t)\right\}$ of this auxiliary
dynamics.
The probability density of seeing $\overline{\Gamma} \equiv \{\overline{m}(t)\}$ in this auxiliary dynamics is
\bea
\overline{P} \left[ \overline{\Gamma} \right]=\overline{p}_i(\overline{m}_0) \prod_{j=1}^{J}
\exp\Big[-\int_{\tau_j}^{\tau_{j+1}} dt~ \overline{K}_{\overline{m}_j}(t)-\delta_{r\overline{m}_j}\int_{\tau_j}^{\tau_{j+1}} dt
~R^{\text{aux}}_r(t) \Big]
\prod_{j \in J_d}~\overline{W}_{\overline{m}_{j}\overline{m}_{j+1}}(t_f-\tau_j)~\prod_{j \in J_r}~\overline{R}_{\overline{m}_jr}(t_f-\tau_j).
\l{time-reversed-trajectory}
\eea
where $\overline{p}_i$ is the initial condition for the time reversed trajectory.
Unlike in Eq.\eref{forward-trajectory}, the second term in the exponential of Eq.\eref{time-reversed-trajectory} gives us contributions
only when the actual time reversed state $\overline{m}(t)$ is at the reset site $r$.

Crucially, there is a one-to-one mapping between realizations of the resetting dynamics and
their time-reversed counterparts in the auxiliary dynamics. The only requirement for this one-to-one
mapping is that all the resetting transitions are replaced by the \textit{anti-resetting} ones. This leaves some freedom
in choosing the magnitude of various rates in the auxiliary dynamics. To proceed we choose
\bea
\overline{W}_{mn}(t)&=&W_{mn}(t) \nn \\
\overline{R}_{mr}(t)&=&R_{rm}(t)/f(r,m,t),
\l{transition-laws}
\eea
where $f(r,m,t)$ is an arbitrary function.
Namely, we elect to keep the bidirectional transition rates as they were in the resetting dynamics, but allow
for more general choice of the \textit{anti-resetting rates}. We will see in the following that
two specific choices of those rates have an interesting physical interpretation.

The probability density of $\overline{\Gamma}$ can now be rewritten as
\bea
\overline{P} \left[ \overline{\Gamma} \right]=\overline{p}_i(m_J) \prod_{j=1}^{J}
\exp\Big[-\int_{\tau_j}^{\tau_{j+1}} dt~ K_{m_j}(t)-\delta_{r m_j}\int_{\tau_j}^{\tau_{j+1}} dt ~R_r^{\text{aux}}(t) \Big]
\prod_{j \in J_d}~W_{m_{j}m_{j+1}}(\tau_j)~\prod_{j \in J_r}~\frac{R_{r m_j}(\tau_j)}{f(r,m_j,\tau_j)},
\l{time-reversed-fictitious-trajectory}
\eea
where $R_r^{\text{aux}}(t)=\sum_{m \neq r} R_{rm}(t)/f(r,m,t)$. The one-to-one
mapping between the realizations of the resetting and auxiliary dynamics results in an integral fluctuation
theorem. Let us define
\bea
\Sigma \left[ \Gamma \right]\equiv \ln \frac{P \left[ \Gamma \right]}{\overline{P} \left[ \overline{\Gamma} \right]}.
\l{Sigma-definition}
\eea
The fact that the auxiliary dynamics conserves probability means that
\bea
\left\langle e^{-\Sigma}  \right\rangle~=~\sum_{\Gamma}~e^{-\Sigma \left[ \Gamma \right]}~P \left[ \Gamma \right]=
\sum_{\overline{\Gamma}}~\overline{P} \left[ \overline{\Gamma} \right]~=~1,
\l{IFT-discrete-resetting}
\eea
where the average is over an ensemble of realizations of the resetting dynamics.
The Jensen's inequality can now be used to derive a second-law-like inequality, resulting in
$\big\langle \Sigma \left[\Gamma \right] \big\rangle \geqslant 0$.
We note that Eq.\eref{IFT-discrete-resetting} is valid for any choice of the initial condition of the auxiliary dynamics.
We choose the initial distribution of the auxiliary dynamics to be identical to
the final distribution of the resetting dynamics
such that $\overline{p}_i=p_f$.

The functional appearing in the integral fluctuation theorem (\ref{IFT-discrete-resetting}) is given by
\bea
\Sigma[\Gamma]= \Delta S_{\text{tot}}-\Delta S_{\text{reset}}+\Sigma_{\text{dyn}}.
\l{Sigma-defn}
\eea
Here
\bea
 \Delta S_{\text{tot}}=\ln \frac{p_i(m_0)}{p_f(m_J)}+\sum_{j \in J_d}~\ln \frac{W_{m_{j+1}m_j}(\tau_{j+1})}{W_{m_{j}m_{j+1}}(\tau_{j+1})},
 \l{tot-entropy}
\eea
is the total entropy production in the system.
The contributions in Eq.\eref{tot-entropy} are due to the changes in the fluctuating system entropy
and the medium entropy from all
the {\em bidirectional} transitions respectively. The resetting transitions are responsible for
a resetting entropy production term. It has the following form
\bea
\Delta S_{\text{reset}} &=-& \sum_{j \in J_r}~\ln f(r,m_j,\tau_{j+1}).
\l{reset-entropy}
\eea
While the first two terms in $\Sigma[\Gamma]$ have a thermodynamical interpretation,
the last term $\Sigma_{\text{dyn}}$ is dynamical in nature. This dynamical term is given by
\bea
\Sigma_{\text{dyn}}~=~\sum_{j=1}^{J}\delta_{rm_j} \int_{\tau_j}^{\tau_{j+1}} dt~ R_r^{\text{aux}}(t)- \sum_{j=1}^{J}(1-\delta_{rm_j})
\int_{\tau_j}^{\tau_{j+1}} dt~R_{rm_j}(t).
\l{Delta-S-dyn-1}
\eea
This term can be rewritten as
\bea
\Sigma_{\text{dyn}}~=~ \int dt \left[R_r^{\text{aux}}(t) \chi_r (t) - \sum_{m \ne r} R_{rm}(t) \chi_m (t) \right],
\l{Delta-S-dyn-2}
\eea
where $\chi_i(\Gamma)$ is an indicator function so that $\chi_i~=~1$, for $\Gamma(t)=i$ and $0$ otherwise.
In particular,
for autonomous processes  $\Sigma_{\text{dyn}}$ has the following form
\bea
\Sigma_{\text{dyn}}~= \sum_{m \neq r}~ [R_r^{\text{aux}}~ \Theta_r(\Gamma)-R_{rm} ~\Theta_m(\Gamma)],
\l{Delta-S-dyn-3}
\eea
where
$ \Theta_i(\Gamma) \equiv \int~ dt ~ \chi_i(\Gamma)$
is the so-called residence time. It is simply
 the total
time spent at the site $i$ during the realization. The residence time is a stochastic quantity which
fluctuates from one realization to another. When normalized by the observation time, this
quantity is often known as the empirical density, which converges to the steady state distribution of the system.
It is worth emphasizing that Eq.\eref{IFT-discrete-resetting} holds for any
choice of the function $f(r,m,t)$.

Up to now the physical interpretation of $\Sigma \left[ \Gamma \right]$
was not fully clear as it depended on parameters of the non-physical auxiliary process. In the following, we consider
two particular choices of $f(r,m,t)$. These choices
lead to integral fluctuation theorems with interesting physical interpretation that is expressed only in terms of the original
resetting dynamics.

\subsection{$f(r,m,t)~=~1$}
\label{subsec:A}

In this case the auxiliary dynamics is obtained by simply reversing
the direction of the resetting transitions while maintaining
their magnitude. This prescription was previously used to study
Markov processes with unidirectional transitions \cite{Rahav2014}.
Systems with resetting are a subtype of such processes where all the unidirectional
transitions point to one preselected site.

The choice of $f(r,m,t)=1$ is useful since it results in a functional with a physically meaningful
interpretation. The resetting entropy contribution to $\Sigma \left[ \Gamma \right]$ identically vanishes.
In contrast, the dynamical contribution does not. For time-independent transitions we find
\bea
\Sigma_{\text{dyn}}= \sum_{ m \neq r} ~ R_{rm} \left[ \Theta_r(\Gamma)- ~\Theta_m(\Gamma) \right],
\l{sigma-dynamic-independent}
\eea
where we have used the fact that the total absorption rate at the reset site is given by
$R_r^{\text{aux}}=\sum_{m \neq r} R_{rm}$.
The dynamical contribution to $\Sigma \left[ \Gamma \right]$ therefore depends on the fluctuating residence
times at all sites.

The structure of the dynamical term
$\Sigma_{\text{dyn}}$ exhibits similarity to the so-called dynamical activity or traffic which basically
counts the number of all jumps irrespectively of their direction in a general jump process \cite{Maes2008,Baiesi2009}.
An important distinction, however, is that the traffic is time symmetric by construction,
namely it does not not change sign if the trajectories
are observed backward in time. Detailed and integral fluctuation relations for the traffic functional
were derived in \cite{Baiesi2015} based on an artificial auxiliary dynamics \cite{Maes2008,Rahav2014}.

\subsection{$f(r,m,t)~=~\frac{p(r,t)}{p(m,t)}$}

An alternative choice of the auxiliary dynamics is obtained by choosing $f(r,m,t)=p(r,t)/p(m,t)$,
where $p(m,t)$ is the time dependent solution of the master equation. This choice also
results in an integral fluctuation theorem with an appealing physical interpretation. We first notice that
the resetting entropy along a realization is readily obtained from Eq.\eref{reset-entropy}
\begin{equation}
  \Delta S_{\text{reset}}=\sum_{j \in J_r} \ln \frac{p(m_j,t)}{p(r,t)}.
\end{equation}
It is evident that the resetting entropy does not vanish for this choice of resetting rates. Furthermore,
the mean rate of resetting entropy production is given by
\bea
\dot{S}_{\text{reset}}(t) &=& \sum_{m \neq r} ~
R_{rm}(t)~p(m,t)~\ln \frac{p(m,t)}{p(r,t)}.
\l{reset-EP}
\eea
This is precisely the expression derived by Fuchs et. al. \cite{Seifert2016} (see Eq. (24) there).

For this choice of $f(r,m,t)$ the dynamical part of $\Sigma \left[ \Gamma \right]$ is given by
\begin{equation}\label{eq:dyn2choice}
  \Sigma_{\text{dyn}}=\int dt \sum_{m \ne r} R_{rm} (t) \left[\frac{p(m,t)}{p(r,t)} \chi_r (t) - \chi_m (t) \right].
\end{equation}
The appealing feature of this choice of auxiliary dynamics is that the ensemble average of this dynamical
term vanishes. Indeed, by definition $\left< \chi_m (t)\right>=p(m,t)$.
Consequently, the ensemble average of Eq.\eref{eq:dyn2choice} gives us
\begin{equation*}
 \left< \Sigma_{\text{dyn}}\right> =\int dt \sum_{m \ne r} R_{rm} (t) \left[\frac{p(m,t)}{p(r,t)} p(r,t) - p(m,t) \right]=0.
\end{equation*}

The
vanishing mean of the dynamical contribution to $\Sigma \left[ \Gamma \right]$ results in a purely
thermodynamic second-law-like inequality
\bea
\dot{S}_{\text{tot}}-\dot{S}_{\text{reset}}\geqslant 0.
\eea
which is derived by using Jensen's inequality in the integral fluctuation theorem in Eq.\eref{IFT-discrete-resetting}.
At steady state $\dot{S}_{\text{sys}}=0$, and the inequality simplifies to
\bea
\dot{S}_{\text{med}}-\dot{S}_{\text{reset}}\geqslant 0.
\eea
Here $\dot{S}_{\text{med}}$ is the medium entropy production rate.
This version of the second law, applicable to resetting systems, was originally derived in \cite{Seifert2016}.
Our results show that it can also be derived from an integral fluctuation theorem by using the Jensen inequality.
Interestingly, the fluctuating functional appearing in this IFT includes both thermodynamic and dynamical contributions.
The latter turns out to have a vanishing mean
and therefore does not appear in the second law,
but it certainly contributes to the stochastic thermodynamics of the system.

Several recent papers have derived integral fluctuation theorems for systems where only some of the transitions can be
observed \cite{Haritch2014,Shiraishi2015,Polettini,Bisker}.
The derivation is based on a construction of auxiliary dynamics very similar to the one employed here.
It is interesting to note that this particular prescription was found
to be useful in a variety of physical contexts.


\begin{figure}[t]
\centering
\begin{subfigure}{.5\textwidth}
  \centering
  \includegraphics[width=1.3\linewidth]{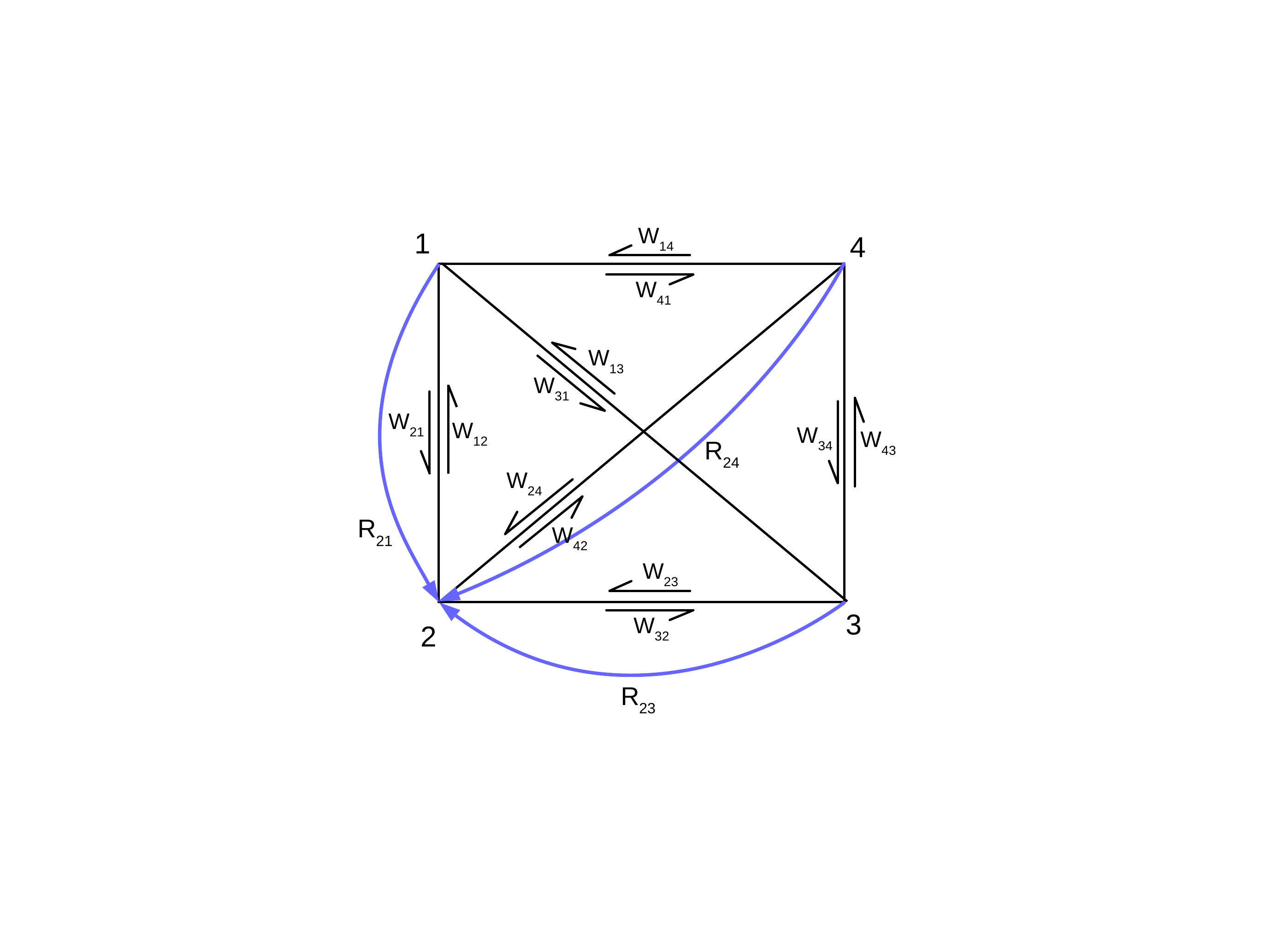}
\end{subfigure}%
\begin{subfigure}{.5\textwidth}
  \centering
  \includegraphics[width=.7\linewidth]{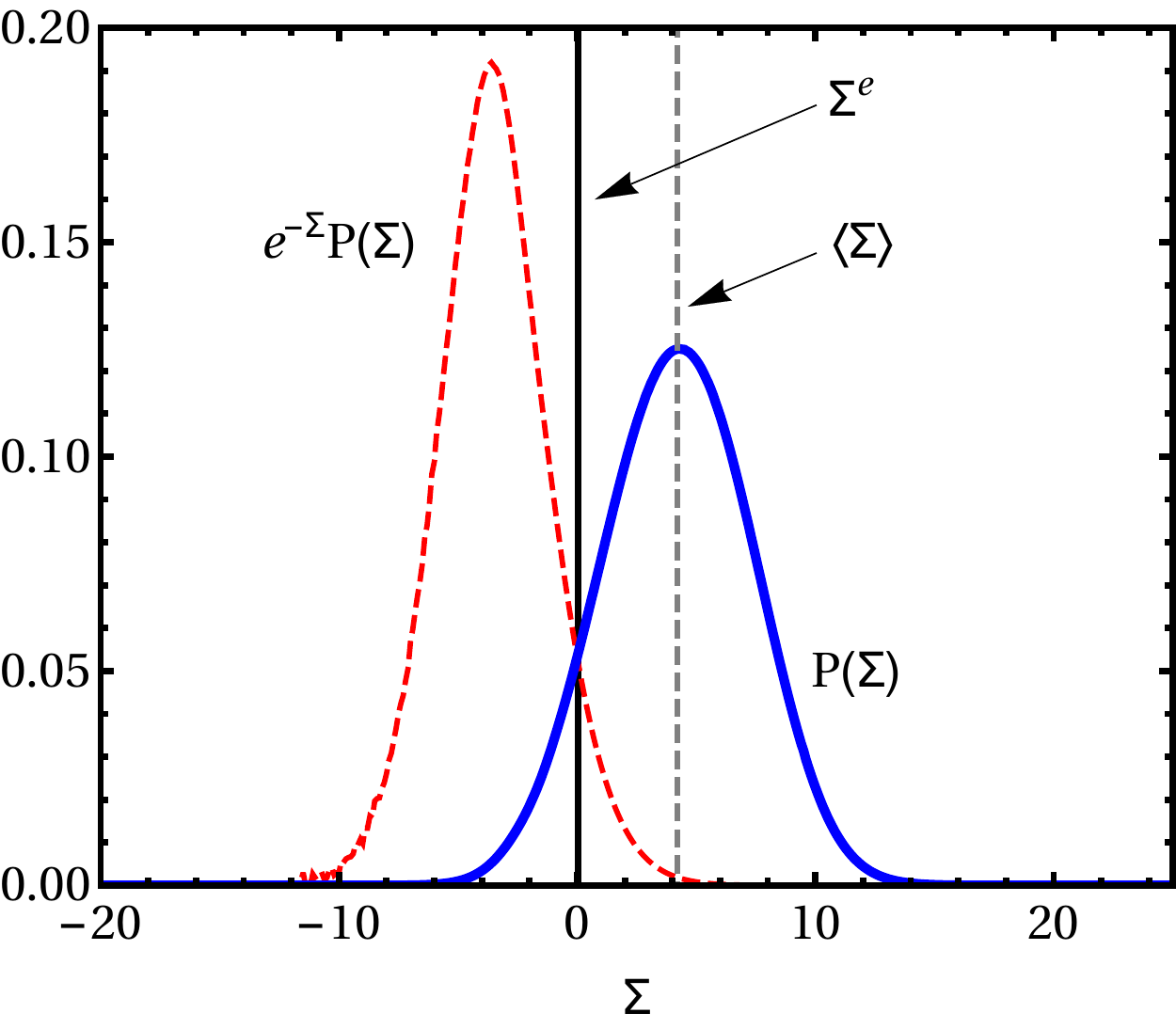}
\end{subfigure}
\caption{\textbf{Left panel}: Schematic representation of a 4-site Markov jump process with resetting to site $2$.
The solid edges (in black) between two states represent the bidirectional transitions. Resetting transitions
(with respective rates) to the site 2 are indicated by
the solid arrows (in blue).
\textbf{Right panel}: Numerical verification of the integral fluctuation theorem for the four-state Markov jump process (depicted in the left panel).
$N_r=10^9$ realizations were used to generate the probability distribution of $\Sigma$. We used the prescription
of Sec. \ref{subsec:A} to compute the functional.
The dashed red line depicts $e^{-\Sigma}P(\Sigma)$ calculated from $P(\Sigma)$. The intersection of these two graphs is at the
origin as expected from the fluctuation theorem. The average of this functional, $\langle \Sigma \rangle$ is always positive and
indicated by grey dashed line. The solid line (in black) depicts
$\Sigma^e = \ln~\frac{1}{N_r}\sum_{i=1}^{N_r}  e^{-\Sigma_i}$, computed by summing over all the realizations.
This value turns out to lie close to the origin ($\Sigma^e=0.04934...$ in this case),
as predicted by the IFT Eq.\eref{IFT-discrete-resetting}.}
\l{fig-unidirection}
\end{figure}


\subsection{Numerical Simulations}

To illustrate our considerations we simulated a stochastic jump process with both bidirectional and resetting transitions.
We used a four-site system as depicted in \fref{fig-unidirection}. While bidirectional transitions can occur between
any two sites,  the resetting transitions can occur only to a preselected site, chosen here to be site $2$.
In our simulation we used the transition rates $W_{12}=0.3, W_{13}=1.0, W_{14}=0.7, W_{21}=0.5, W_{23}=0.6, W_{24}=0.7
, W_{31}=0.9, W_{32}=1.3, W_{34}=0.7, W_{41}=0.8, W_{42}=0.2, W_{43}=1.3, R_{21}=0.4, R_{23}=0.6, \text{and}~R_{24}=1.0$.
We chose an example with time-independent rates to allow usage of the Gillespie algorithm. For this dynamics
the waiting time between realizations is distributed exponentially.
The system is initially in a uniform probability distribution, with $p_i(n) \equiv p(n,0)=1/4$. Then time evolution of $p(n,t)$ is obtained by
solving the master equation.

To explore the stochastic thermodynamical properties of this model we need to generate single realizations of the jump process.
To this aid,
we have used Gillespie algorithm to
generate the stochastic trajectories of the system by determining the epochs of jumps between the states.
We simulated the jump process, by picking an initial state with probability $p_i (n)$,
and then following the transitions that the system makes until a final time of $t_f=5$.
For each realization we computed the functional
$\Sigma$ [Eq.\eref{Sigma-defn}],
using the prescription $f(r,m,t)=1$, which results in a vanishing resetting
entropy production.
The system and the medium entropy are calculated with the help of Eq.\eref{tot-entropy}. This requires following the initial and final states of
each realization, as well as the transitions made during it. The calculation of the system's entropy also requires knowledge of
the initial and final probability distributions.
Finally, the dynamical contribution $\Sigma_{\text{dyn}}$
is computed using Eq.\eref{sigma-dynamic-independent} where the residence time $\Theta_i$ at site $i$ is computed from the stochastic
dynamics simulation.

We obtain the complete statistics of $\Sigma$ by taking
an ensemble over $N_r=10^9$ independent realizations. To test the validity of the integral fluctuation theorem we
computed $\Sigma^e \equiv \ln \frac{1}{N_r} \sum_{i=1}^{N_r} e^{-\Sigma_i}$ numerically. For the given set of parameters, we find
$\Sigma^e=0.04934..$, which is marked by a solid vertical line in \fref{fig-unidirection} (right panel). The dashed vertical line
depicts $\langle \Sigma \rangle$.
$\Sigma^e$ is considerably smaller than $\langle \Sigma \rangle$ and is located near the origin. This is consistent with the 
predictions of the IFT Eq.\eref{IFT-discrete-resetting}. We note in passing that trying to numerically verify the IFT 
Eq.\eref{IFT-discrete-resetting} for processes of longer duration
may be difficult. The reason is the growth of $\langle \Sigma \rangle$ with the duration, resulting in the need to sample
an exceedingly large number of realizations to ensure convergence of the exponential average $\langle e^{- \Sigma} \rangle$.



\section{Conclusion}
\label{Conclusion}

In this paper we have studied the stochastic thermodynamics of resetting systems. In particular, we
investigated whether stochastic dynamics with resetting satisfies fluctuation theorems. Our results are complementary to the ones
recently presented by Fuchs et. al. \cite{Seifert2016}, where the work and entropy production of resetting were identified, and a version of the
second law that holds with resetting was derived.

The search for fluctuation theorems of resetting systems is complicated by the fact that resetting events violate microreversibility.
This violation of microreversibility ultimately means that many of the known versions of fluctuation theorems are inapplicable in
systems with resetting. Nevertheless, we identify two integral fluctuation theorems that hold for stochastic dynamics with resetting.

The first IFT is the Hatano-Sasa relation for transitions between steady states \cite{HatanoSasa2001}. The functional that appears
in the fluctuation relation in systems with resetting includes both the usual excess heat but also a contribution due to resetting entropy change.
Interestingly, while none of these terms behaves like a proper excess quantity, in the sense of exhibiting a small mean for quasistatic processes,
their sum does.

The second IFT describes stochastic jump processes with resetting. It is derived by comparing the resetting dynamics to an auxiliary dynamics
in which resetting is replaced with anti-resetting. There is some freedom in choosing the transition rates of this auxiliary dynamics, resulting
in some freedom in the final form of the IFT. We identify two choices for the auxiliary dynamics which lead to an IFT with interesting physical interpretation.
The first choice leads to an IFT for a functional that has no contribution of resetting entropy.
Instead, it includes a dynamical term that is calculated
from fluctuating residence times in sites.
The second choice leads to a functional that does have a contribution
from the entropy changes in the resetting events. This functional also includes a modified dynamical
contributions, but we find that the ensemble average of this term vanishes. Interestingly,
we find that this fluctuation relation leads to the second-law-like inequality found in Ref. \cite{Seifert2016}.

The Hatano-Sasa relation holds for quite general processes in which a system is driven out
of a steady state. It is therefore very desirable to calculate analytically the distribution
of values of the Hatano-Sasa functional. Several recent papers employed the theory
of large deviations to calculate
 distributions of thermodynamic observable (like work, heat and total entropy production) \cite{VanZon-Cohen, PalSanjib, Touchette}.
 A similar approach
could also be handy to describe the Hatano-Sasa functional.
Another possible research direction would be to study the full statistics of the residence times introduced in Sec. \ref{sec:IFTDR}. This could be done
using the Feynman-Kac formalism following references \cite{SatyaCurrentScience, Sanjib-SatyalLT,PCB2017}, where the authors have studied the
full statistics of the residence time in generic diffusion processes.
The natural extension of these studies to systems with
resetting is to focus on the statistics of the residence time near the resetting point (or state).

\begin{acknowledgments}
This work was supported by the the U.S.-Israel Binational Science
Foundation (Grant No. 2014405), by the Israel Science Foundation (Grant
No. 1526/15), and by the Henri Gutwirth Fund for the Promotion of
Research at the Technion.
\end{acknowledgments}

\end{document}